\documentclass{llncs}
\usepackage{epsfig,endnotes,citesort,latexsym,times,graphicx,subfigure,color,colordvi,amssymb,algorithm2e} 

\newcommand{\affa}{$^{1}$}
\newcommand{\affb}{$^{2}$}
\newcommand{\affc}{$^{3}$}
\newcommand{\affd}{$^{4}$}
\newcommand{\affe}{$^{5}$}

\def\T{\ensuremath{\mathcal{T}}}

\newcommand{\eat}[1]{}
\newcommand{\captionspace}{-10pt}

\makeatletter
\renewcommand\section{\@startsection {section}{1}{\z@}%
   {-1ex \@plus -1ex \@minus -.2ex}%
   {1ex \@plus.2ex}%
   {\normalfont\large\sffamily\bfseries}}
\renewcommand\paragraph{\@startsection{subsubsection}{4}{\z@}%
   {1.5ex\@plus 1ex \@minus .2ex}%
   {-1.5ex \@plus .2ex}%
   {\normalfont\normalsize\sffamily\bfseries}}
\makeatother

\title{Providing Scalable Data Services in Ubiquitous Networks}

\author{%
{Tanu Malik*\affa, Raghvendra Prasad\affb, Sanket Patil\affc, \\ Amitabh Chaudhary\affd, Venkat Venkatasubramanian\affe}
}

\institute{
\affa\hspace{0.12cm}Cyber Center, \affb\hspace{0.12cm}Dept. of Computer Science, \affe\hspace{0.12cm}School of Chemical Engineering \\ Purdue University, USA \\
\affc\hspace{0.12cm}Dept. of Computer Science \\ IIIT, Bangalore, India \\
\affd\hspace{0.12cm}Dept. of Computer Science and Enggineering \\ University of Notre Dame, USA \\
\emph{Contact Author:} tmalik@cs.purdue.edu\\
}

\begin{document}
\maketitle

\begin{abstract}
Topology is a fundamental part of a network that governs connectivity between nodes, the amount of data flow and the efficiency of data flow between nodes. In traditional networks, due to physical limitations, topology remains static for the course of the network operation. Ubiquitous data networks (UDNs), alternatively, are more adaptive and can be configured for changes in their topology. 
This flexibility in controlling their topology makes them very appealing and an attractive medium for supporting ``anywhere, any place'' communication. However, it raises the problem of designing a dynamic topology. The dynamic topology design problem is of particular interest to application service providers who need to provide cost-effective data services on a ubiquitous network. 
In this paper we describe algorithms that decide when and how the topology should be reconfigured in response to a change in the data communication requirements of the network. In particular, we describe and compare a greedy algorithm, which is often used for topology reconfiguration, with a non-greedy algorithm based on metrical task systems. Experiments show the algorithm based on metrical task system has comparable performance to the greedy algorithm at a much lower reconfiguration cost.
\end{abstract}

\section{Introduction}

In the vision of pervasive computing, users will exchange information and control their environments from anywhere using various 
wireline/wireless networks and computing devices \cite{saha2003pervasive}. Although such a definition of pervasive computing is very appealing to users, it has been reported that the technological path for building such an anytime, anywhere networking environment is less clear \cite{das}. A primary technical issue is the configuration of the topology between nodes and devices \cite{saha2003networking}, \cite{das}. Traditional computing environments such as the Internet, the native routing infrastructure is fixed and the topology is predominantly static. However in large-scale pervasive computing environments, the topology is mostly deployed and maintained by application service providers (ASPs) who contract with underlying ISPs and buy network bandwidth between nodes to provide value-added network services to end-systems. Thus they can maintain a more dynamic and flexible topology.

We are interested in understanding how and when  a topology should be configured in order to support distributed data management services. These services can be in the form of replica or a caching service provided by an ASP in which nodes (both wired and wireless) acquire and disseminate data. In a pervasive computing environment, configuring a topology to support such a distributed data service can be challenging. Firstly, nodes and devices have limited resources \cite{saha2003pervasive}. The resource limitation is often due to restricted buffer sizes and storage capacities at nodes, limited bandwidth availability between nodes or limited number of network connections that a node can support. An optimal usage of limited resources requires a topology design that routes the flow of data through most cost-efficient paths. Secondly, the data communication pattern of clients may change drastically over time \cite{saha2003pervasive}, \cite{saha2003networking}. This may require a topology that quickly adapts to the change in data requirements.

While the flexibility of reconfiguring a topology as data communication patterns change is essential, reconfiguring a topology every time a communication pattern changes may not be beneficial. Changing network topology is not cost-free: it incurs both management overhead as well as potential disruption of end-to-end flows. Additionally, data in transit may get lost, delayed, or erroneously routed. In the presence of these costs, it might be useful to monitor changes in the communication pattern at for every query request but the topology should only be changed if the long term benefits of making a change justifies the cost of the change.
%
%

In this paper, we are most interested in the dynamic topology design problem, \emph{i.e.,} the problem of determining how and
when to reconfigure a topology in a resource-constrained pervasive computing environment in which data communication patterns change dynamically. We describe this problem in Section \ref{sec:problem}. In Sections \ref{sec:static} and \ref{sec:reconf} respectively, we describe the operational costs of satisfying a demand pattern in a given topology and the cost of reconfiguring between two topologies. To calculate operational costs on a graph, we describe a linear-programming based solution. 

In Section \ref{sec:dynamic}, we describe the algorithms that decides when and how to reconfigure. We first describe a greedy algorithm, which is a natural, first-order algorithm that one would choose for deciding when to reconfigure. We then describe a non-greedy algorithm based on metrical task systems \cite{borodin}. Task systems are general systems that capture the cost of
reconfiguration between two states in addition to the cost of
satisfying a given demand in a given state. By including the reconfiguration cost the system prevents
oscillations into states that are sub-optimal in the long run. The distinguishing part is that task system based algorithms
require no statistical modeling or aggregation of the communication requirements from the service provider.
This lack of making any assumptions about how the communication requirements may change over time allows the algorithm to provide
a minimum level of guarantee of adapting to changes in the communication requirement.
Greedy algorithms on the other hand are heuristic and provide no theoretical evidence or a systematic way of understanding when to change a topology. We evaluate the performance of all algorithms in Section \ref{sec:expt} and conclude in Section \ref{sec:conclusion}.

\vspace{-5pt}
\section{Related Work}
\label{sec:related}

The topology design problem has received significant interest in large information systems such as 
optical networks, data-centric peer-to-peer networks, and more recently complex networks. 
In these systems the objective is to design an optimal topology under arbitrary optimality requirements of efficiency, cost, balance of load on the servers and robustness to failures. The design of an optimal topology is obtained by deriving these measures from past usage patterns and then using network simulation to obtain an optimal topology.
Such simulations, extensively described in \cite{patil-dht}, \cite{patil-breeding}, are often based on neural networks and genetic algorithms and in which optimal topologies are obtained after executing the software for several hours. The premise is that once an optimal topology is chosen then it will remain static for the duration of the network operation. 

In most adaptive networks such as ubiquitous networks \cite{saha2003pervasive} and overlay networks \cite{shenkar}, communication pattens vary so significantly that it is often difficult to obtain an representative usage pattern to perform a simulation. In the past \cite{patil-breeding}, research proposed to perform simulation repeatedly to obtain an optimal topology and the system reconfigures itself. However, in these systems communication patterns are aggregated over large time scales and the reconfiguration is slow. Recently \cite{mcauley2000self} adaptive networks have focused on auto-configuration \cite{mcauley2001experience} in which systems self-monitor the communication requirements and reconfigure the topology when communication patterns change drastically. The dynamic topology problem has been recently studied in the context of overlay networks in which topologies can be designed either in the favor of native networks, which improves performance
or the ultimate customers which reduces the operation cost of the whole system \cite{fan}. While our problem is similar to theirs, our context and cost metrics are different: We study the dynamic topology problem in the context of ubiquitous environment in which nodes and devices have limited resources and communication requirements change arbitrarily.

Given an optimal topology and a stable communication pattern, the problem of determining how to use the edges of the topology 
such that the cost of using the topology is minimized is itself an intractable problem. Several versions of the problem  have been studied in computer networks under the class of multi-commodity flow problems \cite{baruch}, \cite{fan}. In this paper, for simplicity, we have restricted ourselves to single commodity flows \cite{ortega} which is a suitable model when considering communication requirements over a set of replica nodes. Our primary focus is to understand how to adaptively move between optimal topologies when communication patterns change dynamically.

 
\section{Minimizing the Cost of Data Sharing in a UDN}
\label{sec:problem}

We consider a ubiquitous computing environment established by an application service provider to provide data sharing services across the network. The provider, strategically, places replicas on the network to disseminate data. Clients (which can be wired or wireless) make connections to one of the replicas and send queries to it. The query results in a variable amount of data being transfered from the replica to the client. Query results are routed according to the topology of the network. The application service provider pays for the usage of the network, i.e., the total amount of data that passes through the network per unit of time. 
The service provider would like to use those edges of the topology through which the cost of transferring the data is minimized
subject to data flow constraints over the network. We now state the problem formally. 

Let the topology, $T$, be represented by a graph $G = (V,E)$ in which $V$ denotes the set of all nodes in the network and $E$ denotes the set of all edges. Let there be $P$ replicas and $C$ clients on the network such that $P \cup C \subseteq V$ and $P \cap C = \emptyset$. Each edge $e \in E$ in the topology $T$ has a cost $c_e$ which is the cost of sending unit data (1 byte) through each pair of nodes. Let $b_e$ denote the maximum amount of bytes that can be sent on any edge. Each client, $C_i$ receives an online sequence of queries $\sigma_{C_i} = (q_1, \ldots, q_n)$. The data requirement of each query, $q_i$, is assumed equivalent to its result size.
We denote the data communication requirements of all clients 
by 
$\sigma = (\sigma_{C_i}, \ldots, \sigma_{C_M})$.

A topology $T$ is chosen from a feasible set of topologies. Given $\left|V\right|$ nodes, theoretically, there are 
a total of $2^{\left|V\right|(\left|V\right|-1)/2}$ possible topologies.
However, not all of these topologies are desirable in practice. 
A topology is usually required to be connected so
that every node remains in contact with the rest of the 
network. In addition a topology may be either symmetric (regular graph) or scale-free \cite{barabasi2003scale}. In a symmetric topology nodes have nearly identical degree distributions and share uniform load. In scale-free topologies  
some of the nodes act as ``super nodes'' and have a relatively larger load than other
nodes. Scale-free topologies have increasingly been shown to be a better design choice for 
peer-to-peer data networks \cite{patil-breeding}, \cite{patil-dht}.
To take into account the effect of symmetric and scale-free topologies, we assign a factor, $\rho \in [0,1]$, for a topology which measures the skew in degree distribution \cite{patil-breeding}. $\rho$ is defined
as:
\begin{equation}
\rho = 1 - \frac{\left|V\right|(\hat{p} - \bar{p})}{(\left|V\right|-1)(\left|V\right|-2)}
\end{equation}
in which $\hat{p}$ is the maximum degree in the graph and $\bar{p}$ is the average degree of the graph. 
Thus $\rho$ is 0 for a scale-free topology such as a star and 1 for a symmetric topology such as a circle. 
We denote the set of all feasible topologies, which have a given $\rho$, by 0-1 adjacency matrices $\mathcal{T_{\rho}} = {T_1, T_2, \ldots, T_N}$.

Finally, a topology reconfiguration algorithm is the sequence of topologies $T = $ \\ $(T_1, \ldots, T_n), T_i \in \mathcal{T}$ 
used by the UDN over time in response to the communication requirement, 
$\sigma$, changing over time. The total cost is defined as
\begin{equation}
cost(\sigma,T) =  \sum_{i=1}^{n}\sigma(T_i) + \sum_{i=0}^{n-1} d(T_i,T_{i+1}),
\end{equation}
in which the first term is the sum of costs of satisfying data requirement of all clients $\sigma$ under the
corresponding topology and the second term is the sum of costs of transitioning
between topologies in $T$. In the first term costs under a given topology are calculated under the assumption that data communication pattern remains static. The second term is the cost of reconfiguring a topology. Note, if $T_{i+1} = T_i$ there is no real change
in the topology schedule and incurred reconfiguration cost is zero.

The total cost equation is minimized by an algorithm which generates the best topology schedule $T$. 
This requires an algorithm to identify when demand characteristics have changed significantly such that the 
current physical design is no longer optimal and choosing a new topology such that excessive costs
are not incurred in moving from the current topology, relative to the benefit. 
An offline algorithm that knows the entire $\sigma$ obtains a configuration schedule $S$ with the minimum cost and is optimal.
An adaptive algorithm, determines $T=(T_0, . . . , T_n)$ without seeing the complete workload $\sigma=(q_1, . . . , q_n)$ and works in an online fashion. We first describe the cost estimation functions which can be used by any algorithm and then describe algorithms which decide when and how to configure.


\subsection{The Topology Problem under Static Communication Pattern}
\label{sec:static}

When the data communication pattern is static, the system will remain in that topology that minimizes the cost of satisfying the 
pattern. We describe a linear-programming based solution for measuring the cost of satisfying a data communication pattern over a given topology. Given an edge in a topology $T$, recall that the cost of flowing a unit amount of data through that  edge is 
$c_e$ and the maximum amount of bytes that can be sent on any edge is $b_e$. If $f_e$ is the amount of bytes that flow through this edge in order to satisfy the communication requirement at a client, then the overall cost to support communication requirement of all clients is the cost of flowing data through all the edges which is defined as
\begin{equation}
\label{eq:0}
\sum_{e \in E} \left|f_e\right|. c_e
\end{equation}
This cost must be minimized to subject to the following constraints:
\begin{itemize}
\item The flow in an edge should not exceed its capacity and there is no excess reverse flow in an edge.
\begin{equation}
\label{eq:1}
\forall e = (u,v) \in E: f_{u,v} = - f_{v,u} \textit{ and } \left|f_e\right| \le b_e
\end{equation}
\item The replica nodes do not request data. 
\begin{equation}
\label{eq:2}
\forall p \in P, \forall \textit{ u } \in \textit{neighbors of } p: f_{u,p} \le 0
\end{equation}
\item If the communication requirement at each client is static and equals $\sigma_{C_i}$ then the entire requirement is satisfied: 
\begin{equation}
\label{eq:3}
\forall c \in C: \sum_{u \in \textit{ neighbors of } c} f_{u,c} = \sigma_{C_i}
\end{equation}
\item The skew in the flow of data should correspond to the skew in the topology $\rho$. For this, we also restrict the number of bytes passing through each node $b_u, u \in V$. 
\begin{equation}
 \rho =  \textit{max}(b_u) - \frac{\sum_{u} b_u}{\left|V\right|}, 
\end{equation} 
 where
 $\forall u \in V: b_u = \frac{\sum_{v \in neighbor \:of\:u} \left| f_{vu} \right|}{2}$
\end{itemize}
If the data were routed through using minimum-operation-cost paths, the static topology design problem is the problem of finding a topology $T$ , under the constraints of connectivity and degree-bound, that can minimize the cost in Equation \ref{eq:0} for a communication requirement $\sigma$ that remains constant over time. We term such a topology, optimal-static topology for $\sigma$, and denote it by $T^*(\sigma)$. Similar to most other topological design problems, the static topology design problem can be modeled
as a linear programming problem and can be solved efficiently in the worst case.


\subsection{The Reconfiguration Cost}
\label{sec:reconf}

Every time the system reconfigures its topology to adapt to changes in communication requirements, a reconfiguration cost is incurred. This cost is the overhead or the impairment to performance incurred by the transition from one topology to another.
Various costs could be incurred during a topology reconfiguration, depending on the implementation details of the
UDN. For example, establishing and changing links incurs control and management
overhead which can be translated to energy costs in a wireless network or 
costs paid to ISPs in a wired network or a combination of both in wired/wireless setting \cite{mcauley2000self}. 
Any fraction of data in transit during topology reconfiguration is subject to
routing disturbance leading to a rerouting overhead. Depending
on the UDN implementation, when topologies change, data in transit may wander through a path
with a high operation cost. Finally, rerouting overhead can
be magnified at the end-systems.

In this paper, we assume reconfiguration costs as the cost of auto-configuring the entire network. Configuring a network involves 
first establishing basic IP-level parameters such as IP addresses and addresses of key servers
and then automatic distribution of these IP configuration parameters in the entire network \cite{mcauley2001experience}.
In the wired networking environment, protocols such as Dynamic
Host Configuration Protocol (DHCP) \cite{droms1999automated} and Mobile IP \cite{perkins1997mobile} can configure individual hosts. 
In the pervasive environments, Dynamic Configuration Distribution Protocol (DCDP) is a popular protocol for auto-configuration \cite{das}. In DCDP, auto-configuration is done by recursively splitting the address pool down a spanning tree formed out of the graph topology. 
Thus the total configuration cost of the network is essentially proportional to the
height of the spanning tree. 
A general approximate measure for the reconfiguration cost is the 
total number of links that need to be changed during a transition
\begin{equation}
d(T_{old},T_{new}) = \sum W.(g(T_{old}) + g(T_{new})) 
\end{equation}
in which $g(\cdot)$ is the auto-configuration cost and is proportional to the height of the spanning tree in each topology and $W$
is weight parameter that converts this cost in terms of operation costs.

\section{The Topology Reconfiguration Problem with Dynamic Data Requirements}
\label{sec:dynamic}
In a real-world, client nodes receive a sequence, $\sigma$, of queries, in which the size of the query result differs. Thus the amount of data delivered from the replica to the client changes over time. In such a dynamic scenario no one topology remains optimal and a reconfiguration of topology is needed. In this section, we first describe a greedy algorithm which specifies \emph{when} the topology should be reconfigured by looking at the past workload. We consider several variations of this algorithm by considering different lengths of consideration of the past period. 
In several environments, request for data is bursty in that arbitrarily large amounts of data are requested over short periods of time. For such environments, it is difficult to ascertain the length of the past period precisely. We describe a more conservative algorithm based on metrical task systems \cite{borodin}. Algorithms in task systems achieve a minimum level of performance for any workload 
and provide guarantees on the total cost of satisfying data demands and making transitions. 


\vspace{5pt}
{\bf Greedy Algorithm:} Such an algorithm chooses between neighboring topologies greedily. The current topology ranks its neighboring
topologies based on past costs of the communication requirement in the other topologies.
The algorithm keeps track of the cumulative penalty of remaining in the current
topology relative to every other neighboring topology for each incoming data demand.
A transition is made once the algorithm observes that the benefit of a new configuration exceeds a threshold. The threshold is
defined as the sum of the costs of the most recent transition and next
transition that needs to be made. There are various policies of choosing the length of the past interval:


\begin{itemize}
\item {A reactive policy:} The system transitions to another topology every time the cost of satisfying the demand in the current topology is higher than the sum of cost of transitioning and the cost of satisfying the demand in another topology. Thus the system may potentially transition on every input of the demand request. 
\vspace{5pt}

\item {A lazy policy:} The system is slow in transitioning in that it waits for a delta period before deciding to transition to another topology. Thus the system transitions to another topology when the total cost of satisfying demand in a $\delta$ period is lower than the cost of transitioning and satisfying it in the current topology. The reactive and lazy policies are memoryless in that 
the policy does not take into account the past demand patterns.
\vspace{5pt}

\item {An averaging policy:} This policy remembers the demand pattern by considering a demand that is averaged over several $\delta$ periods using a weighting scheme. The system switches to that topology which has the lowest cost of executing the averaged demand. 
\vspace{5pt}
\end{itemize}

{\bf  A conservative algorithm:} Our conservative policy is based on task systems \cite{borodin}. We first consider a task system in which there are only two possible toplogies. A conservative algorithm in such a system works similar to the algorithm for the online ski-rental problem. A skier, who doesn't own skis, needs to decide before every skiing trip that she makes whether she should rent skis for
the trip or buy them. If she decides to buy skis, she will not
have to rent for this or any future trips. Unfortunately, she
doesn't know how many ski trips she will make in future,
if any. 
A well known on-line algorithm for this problem is rent skis as long as the
total paid in rental costs does not match or exceed the purchase
cost, then buy for the next trip. Irrespective of the
number of future trips, the cost incurred by this online algorithm
is at most twice of the cost incurred by the optimal
offline algorithm. If there were only two topologies and the cost function $d(\cdot)$ satisfies symmetry,
the reconfiguration problem will be nearly identical to online ski rental.
Staying in the current topology corresponds to renting skis and
transitioning to another topology corresponds to buying skis.
Since the algorithm can start a ski-rental in any of the states,
it can be argued that this leads to an conservative policy on two states that costs no more than
four times the offline optimal.
%

When there are more than two topologies the key issue is to decide which topology to compare with the current one. This will establish a correspondence with the online ski rental problem. A well-known algorithm for the $N$-state task system is by Borodin et. al. \cite{borodin}. Their algorithm assumes the state space of all topologies to form a metric which allows them to define a \emph{traversal} over $N$ topologies. We show that our reconfiguration function is indeed a metric function and then describe the algorithm.

To form a metric space, the reconfiguration function should satisfy the following properties:
\begin{itemize}
\item $d(T_i, T_j) \ge 0, \forall i \neq j, T_i, T_j \in \mathcal{T}$ (positivity),
\item $d(T_i, T_i) =  0, \forall i \in \mathcal{T}$ (reflexivity), 
\item $d(T_i, T_j) + d(T_j, T_k) \ge d(T_i, T_k), \forall\: T_i, T_j, T_k \in \mathcal{T}$ (triangle inequality), and
\item  $d(T_i, T_j) = d(T_j, T_i), \forall\: T_i, T_j \in \mathcal{T}$ (symmetry).
\end{itemize}
In our case the reconfiguration function $d(\cdot)$ depends upon the sum of reconfiguration costs in the old ($T_i$) and the new topology ($T_j$). Reconfiguration costs are primarily determined by the height of spanning tree over the topologies \ref{sec:reconf} and thus satisfy all the above properties. 

When the costs are symmetrical, Borodin et. al. \cite{borodin} use \emph{components} instead of configurations to
perform an online ski rental. In particular their algorithm
recursively traverses one component until the query execution cost incurred
in that component is approximately that of moving to the other component, moving
to the other component and traversing it (recursively), returning to the
first component (and completing the cycle) and so on.
To determine components, they consider a complete, undirected graph $G'(V,E)$ on $\mathcal{T}$
in which $V$ represents the set of all configurations, $E$ represents
the transitions, and the edge weights are the transition costs.
By fixing a minimum spanning tree (MST) on $G'$, components are recursively determined
by pick the maximum weight edge, say $(u, v)$, in the MST, removing it from the MST.
This partitions all the configurations into two smaller components and the
MST into two smaller trees. The traversal is defined in Algorithm \ref{alg:traversal}.
In \cite{malik09}, the algorithm is shown to have a performance that is atmost $8(N-1)$ worse than the performance of an offline algorithm (one that has complete knowledge of $\sigma$).

\begin{algorithm}[ht]
  \SetLine
  \KwIn{Graph: $G'(V,E)$ with weights corresponding to $d(\cdot)$, Query Sequence: $\sigma$}
  \KwOut{Vertex Sequence to process $\sigma(t)$: $u_0,u_1,\ldots$}
  Let $B(V,E)$ be the graph $G'$ modified s.t.\ $\forall (u,v) \in
  E$ weight $d_B(u,v) \leftarrow d_G'(u,v)$
  rounded to next highest power of 2\;
  Let $F$ be a minimum spanning tree on $B$\;
  $\T \leftarrow \mathsf{traversal}(F)$\;
  $u \leftarrow S_0$\;
  \While{there is a query $q$ to process}
  {
    $c \leftarrow q(u)$\;
    Let $v$ be the node after $u$ in $\T$\;
    \While{$c \geq d_B(u,v)$}
    {
      $c \leftarrow c - d_B(u,v)$\;
      $u \leftarrow v$\;
      $v \leftarrow$ the node after $v$ in $\T$\;
     }
    Process $q$ in $u$\;
  }
  \caption{A Task System-based Algorithm}
  \label{alg:onlinepd}
  \vspace{\captionspace}
\end{algorithm}

\begin{algorithm}[ht]
    \SetLine
    \KwIn{Tree: $F(V,E)$}
    \KwOut{Traversal for $F$: $\T$}
    \uIf{$E = \{\}$} {$\T \leftarrow \{\}$\;}
    \uElseIf{$E = \{(u,v)\}$} {Return $\T$: Start at $u$, traverse to
      $v$, traverse back to $u$\;}
    \Else{
      Let $(u,v)$ be a maximum weight edge in $E$, with weight
      $2^M$\;
      On removing $(u,v)$ let the resulting trees be $F_1(V_1,E_1)$
      and $F_2(V_2,E_2)$, where $u \in V_1$, and $v \in V_2$\;
      Let maximum weight edges in $E_1$ and $E_2$ have weights
      $2^{M_1}$ and $2^{M_2}$\ respectively;
      $\T_1 \leftarrow \mathsf{traversal}(F_1)$\;
      $\T_2 \leftarrow \mathsf{traversal}(F_2)$\;
      Return $\T$: Start at $u$, follow $\T_1$ $2^{M-M_1}$ times, traverse
      $(u,v)$, follow $\T_2$ $2^{M-M_2}$ times, traverse $(v,u)$\;
    }
    \caption{$\mathsf{traversal}(F)$}
    \label{alg:traversal}
      \vspace{\captionspace}
\end{algorithm}

\section{Experiments}
\label{sec:expt}

Our current objective is to get a validation of our policies and algorithms through a simulated environment. Thus while our set-up is a representation of a real-world pervasive environment, doing experiments with real data is part of future work. Our setup simulates a replica environment with 10 replicas and a large number of clients i.e., 40 and 90. Thus the total number of nodes, $\left|V\right|$, is 50 and 100. To determine the feasible set of topologies that are connected and have a given skew in degree distribution, $\rho$, we adopt the following procedure: For a given graph, we fix the maximum node degree $\hat{p}$ and the average degree of the graph, $\bar{p}$. This determines the acceptable skew in the degree distribution. We input different values of $N$, $\hat{p}$, and $\bar{p}$ as parameters to a random graph generator and generate a large number of initial graphs. The generated graphs have no self-loops. In addition, graphs that are disconnected are filtered out as well as graphs that have nodes with less than $\bar{p}$ degree. We assign a cost matrix and an edge capacity matrix with each topology. We choose a set of 100 feasible topologies with a $\rho$ of 0.9. The cost values and the edge capacity values are random values chosen in the range of [100,150] and 
[50,80] respectively.

The clients receive an on-line query sequence in which each query results in $d$ amount of data from the replica. We are not concerned with the actual syntax of the query but the amount of data it generates on the network. We generate the demand sequence at each client from a normal distribution in which the mean changes as a Markov process. In particular, the change in the value of the mean is done after fixed number of time steps and the change in the value is done using a standard exponential moving average.  To model the real world pervasive environment, each client also receives a burst in its demand modeled by a sudden impulse generated randomly for each client. Finally, a sequence of 20,000 events is generated in which data demand at each client is random value in the range [0.1, 100]. 

We use the GAMS \cite{gams} software to solve the static optimal topology problem. GAMS offers an environment to express
mathematical constructs of a linear program. It solves the linear program and  returns the optimal flow of data, $f_{u,v}$, on a topology. We use the optimal flows returned by GAMS to calculate the operation 
costs of satisfying a given set of client demands in a given topology. Reconfigurations costs are calculated by the topology structure, the height of the spanning tree varies from 40-200 and by choosing a $W$ parameter that converts reconfiguration cost in terms of operation costs. We choose $W$ as 200. To perform all these simulations, we developed a Python based system that acts as both an event driven simulator and also a simulator for performing experiments on a UDN. The delta period in the greedy algorithm is chosen to be 50. The MST in the conservative algorithm is implemented using Prim's algorithm.

\subsection{Cost of Reconfiguration}

\begin{figure*}
\centering
\subfigure[N = 50]{
\includegraphics[width=2.25in]{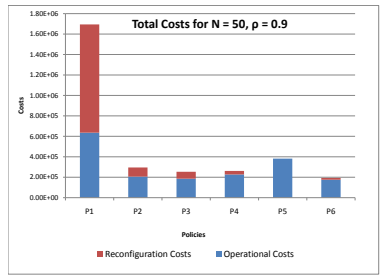}
\label{fg:cost-a}
}
\subfigure[N = 100]{
		\includegraphics[width=2.25in]{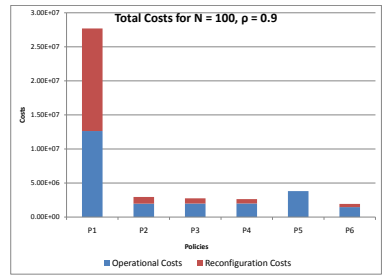}
\label{fg:cost-b}
}

\caption{Total operational and reconfiguration costs}
\end{figure*}

We compute the total cost of satisfying a query sequence under a topology schedule generated by various policies of the greedy algorithm (Policies \texttt{P1-P3}) and the non-greedy algorithm \texttt{P4}. We compare the total cost of adaptive policies with a policy, \texttt{P5}, that does not change its topology at all but remains in a topology that has the minimum operation cost for the entire demand sequence. We also compare with a static optimal policy \texttt{P6} that knows the entire demand sequence in advance. Figure \ref{fg:cost-a} and Figure \ref{fg:cost-b} shows the total cost and its division into operational cost and the cost of transitioning between topologies. 

\texttt{P4} improves on the total cost of \texttt{P5} by 38\%. This is a very encouraging result for pervasive environments
where devices are resource-constrained and policies that improve operation costs are needed. However, \texttt{P3} further improves cost by 42\%. This is because \texttt{P3} relies on the predictive modeling of the demand. However, the
improvement is low considering that \texttt{P4} is general and makes no assumptions 
regarding workload access patterns. The costs of both \texttt{P3} and \texttt{P4} are comparable to the cost of \texttt{P6}. Both \texttt{P1} and \texttt{P2} incur high costs. \texttt{P1} suffers due to being over-reactive making changes even when they are not required and incurs a very high transition cost. \texttt{P2}, by its nature, incurs lower transition costs but high operation costs. On the other hand \texttt{P4} incurs much lower transition costs than \texttt{P3}. This artifact is due to the conservative nature of \texttt{P4}. It evaluates only two alternatives at a time and transitions only if it expects significant
performance advantages. On the other hand, \texttt{P3} responds quicker to workload
changes by evaluating all candidate topologies simultaneously and choosing a topology
that benefits the most recent sequence of queries. This optimism of \texttt{P3}
is tolerable in this workload but can account for significant transition costs in workloads
that change even more rapidly.

\vspace{-15pt}
\subsection{Quality of a Schedule}

\begin{figure*}
\centering
\subfigure[Quality of a Schedule]{
\includegraphics[width=2.25in]{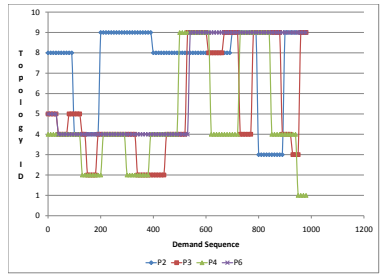}
\label{fg:quality}
}
\subfigure[Effect of Degree Bound]{
		\includegraphics[width=2.25in]{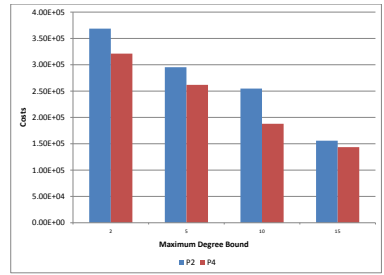}
\label{fg:degree}
}
\caption{Cost Analysis}
\end{figure*}

In this experiment, we compare the quality of a schedule generated over the length of the sequence. We determine the quality of a schedule by comparing the policies with a static optimal policy \texttt{P6} that knows the entire demand sequence in advance.
For presentation sake (Figure \ref{fg:quality}), we omit showing the schedule of \texttt{P1} as it makes so many transitions over the sequence that it affects the presentation of other policies. We also omit \texttt{P5} as it has only one topology in a schedule. We also show the schedule adopted by a policy over 1000 requests as showing over the entire demand sequence suppresses interesting behavior. Results over other demand requests are similar. The experiment is performed over a 50 node topology with W = 2000 and $\rho$ = 0.9. Figure shows that \texttt{P4} closely follows the static optimal policy, which is an artifact of its conservative nature. \texttt{P3} makes lot more transitions than \texttt{P4} because it quickly reacts to changes in workload. It does finally settle on the same states as \texttt{P4}, however at a much higher transition cost. \texttt{P2} does not adapt with the demand sequence and produces a poor schedule.

\vspace{-15pt}
\subsection{Effect of Degree Bound}
\vspace{-8pt}
We are interested in understanding how dynamic topology reconfiguration policies are affected by the degree bound. We have two observations (Figure \ref{fg:degree}): First, the cost of the policies decreases when the degree bound increases. With larger degree bound, there are more feasible topologies and thus the system is able to find better topologies with lower operational costs. Larger degree bound may also decrease reconfiguration costs depending upon the protocol implementation as the system may result in a smaller  height of the spanning tree. While this suggests that in topology design a larger degree bound should be chosen, increasing the degree bound 
shows a decrease in the operational and reconfiguration cost. This is an initial result and we plan to work further on the effect of degree bound on reconfiguration costs.

\vspace{-5pt}
\section{Conclusion}
\label{sec:conclusion}

We have studied the problem of dynamically reconfiguring
the topology of a UDN in response to the changes
in the communication requirements. We have considered two
costs of using the network: the operational cost of transferring data between nodes and the reconfiguration
cost. The objective is to find the optimal
reconfiguration policies that can minimize the potential overall
cost of using the UDN. 
Our policies use both greedy and conservative approaches for adapting to the changes in the communication requirements. We tested the performance of our policies on a medium-size ubiquitous data network and observed shown that dynamic overlay topology reconfiguration can significantly reduce the overall cost of providing a data service over a UDN.

\bibliographystyle{splncs}
\bibliography{udm10}

\end{document}